\documentstyle[aps,twocolumn,10pt]{revtex}
\newlength{\defaultparindent}
\newenvironment{Default Paragraph Font}{}{}
\begin{document}

\title{Two-vibron bound states in $\alpha$-helix proteins : the interplay between the intramolecular anharmonicity and the strong vibron-phonon coupling.}

\author{V. Pouthier}
\address{Laboratoire de Physique Mol\'{e}culaire, UMR CNRS 6624. Facult\'{e} des
Sciences - La Bouloie, \\ Universit\'{e} de Franche-Comt\'{e}, 25030 Besan\c {c}on cedex, 
France.}
\date{\today}
\maketitle

\begin{abstract}

The influence of the intramolecular anharmonicity and the strong vibron-phonon coupling
on the two-vibron dynamics in an $\alpha$-helix protein is studied within a modified Davydov model. The intramolecular anharmonicity of each amide-I vibration is considered and the vibron dynamics is described according to the small polaron approach.
A unitary transformation is performed to remove the intramolecular anharmonicity and a modified Lang-Firsov transformation is applied to renormalize the vibron-phonon interaction. Then, 
a mean field procedure is realized to obtain the dressed anharmonic vibron Hamiltonian.
It is shown that the anharmonicity modifies the vibron-phonon interaction which results in an enhancement of the dressing effect. In addition, both the anharmonicity and the dressing favor the occurrence of two different bound states which the properties strongly depend on the interplay between the anharmonicity and the dressing. Such a dependence was summarized in a phase diagram which characterizes the number and the nature of the bound states as a function of the relevant parameters of the problem. For a significant anharmonicity, the low frequency bound states describe two vibrons trapped onto the same amide-I vibration whereas the high frequency bound states refer to the trapping of the two vibrons onto nearest neighbor amide-I vibrations. PACS: 03.65.Ge, 63.20.Ry, 63.22.+m, 87.15.-v

\end{abstract}

\section{Introduction}

In low-dimensional molecular lattices, the nonlinear nature of vibrational excitons (vibrons) plays a key role for energy transfer as well as energy storage in both physical, chemical and biological systems. In a general way, two main sources yield a nonlinear dynamics, namely the intrinsic intramolecular anharmonicity of each molecule and the extrinsic coupling between the vibrons and surrounding low frequency excitations (such as phonons for instance). 
 
In classical lattices, the anharmonicity gives rise to the occurences of intrinsic localized modes, or discrete breathers, which have been the subject of intense 
researches during the last decade (for a recent review see for instance Refs. \cite{kn:aubry,kn:flach,kn:mackay}). These highly localized vibrations do not require integrability for their existence and correspond to quite general and robust solutions \cite{kn:takeno}. Since discrete breathers favor a local accumulation of the energy which might be pinned in the lattice or may travel through it, they are expected to be of fundamental importance for both energy storage and transport.
Unfortunately, no clear evidence has yet been found for the existence of breathers in real molecular lattices. By contrast, in the quantum regime, two-vibron bound states (TVBS) have been observed in several low-dimensional systems \cite{kn:guyot1,kn:sih,kn:shen,kn:jakob1,kn:jakob2,kn:jakob3,kn:jakob4,kn:jakob_p1,kn:okuyama}.
In that case, the intramolecular anharmonicity breaks the vibron independence and favors the formation of bound states \cite{kn:kimball,kn:bogani,kn:scott,kn:pouthier1,kn:pouthier2,kn:pouthier3}.
When two vibrons are excited, a bound state corresponds to the trapping of two quanta over only a few neighboring molecules with a resulting energy which is lesser than the energy of two quanta lying far apart. The lateral interaction yields a motion of such a state from one molecule to another, thus leading to the occurrence of a delocalized wave packet with a well-defined momentum. As a result, TVBS are the first quantum states which experience the nonlinearity and can thus be viewed as the quantum counterpart of breathers or soliton excitations \cite{kn:scott}.   

The second source of nonlinearity, which originates in the coupling between vibrons and low frequency excitations, was first pointed out by Davydov and co-workers \cite{kn:davydov} to explain the mechanisms for bioenergy transport. 
The main idea is that the energy released by the hydrolysis of ATP can be stored in the 
high frequency C=O vibration (amide-I) of a peptide group of a protein. The dipole-dipole coupling between the different peptide groups leads to the delocalization of the internal vibrations and to the formation of vibrons. However, the coupling 
between the vibrons and the phonons of the protein yields a nonlinear dynamics which counterbalances the dispersion created by the dipole-dipole interaction. It leads to the creation of the so called Davydov's soliton which provides an approximation to the self-trapping phenomena described by a Fr$\ddot{o}$hlich type Hamiltonian \cite{kn:Frohlich}. Soliton mechanisms for bioenergy transfer in proteins have received increasing attention 
during the last twenty five years and a broad review can be found in Refs.\cite{kn:scott1,kn:chris}. 

However, as pointed out by Brown and co-workers \cite{kn:brown1,kn:brown2} and by Ivic and co-workers \cite{kn:ivic1,kn:ivic2,kn:ivic3}, the solution of the Davydov's problem is rather a small vibron-polaron than a vibron-soliton. Indeed, the self-trapping process exhibits two asymptotic solutions depending on the values taken by the three relevant parameters of the problem, i.e. the vibron bandwidth, the phonon cutoff frequency and the small polaron binding energy proportional to the strength of the vibron-phonon coupling. When the vibron bandwidth is greater than the phonon cutoff frequency, the adiabatic limit is reached. The phonons behave in a classical way and create a quasi-static potential well responsible for the trapping of the vibron. The vibron, dressed by this lattice distortion, forms a polaron which is described according to the soliton theory of Davydov. By contrast, when the vibron bandwidth is lesser than the phonon cutoff frequency, the situation corresponds to the non-adiabatic limit in which the quantum behavior of the phonons plays a crucial role. As mentioned by the authors, this situation corresponds to the vibron dynamics in proteins. Therefore, a vibron is dressed by a virtual cloud of phonons which yields a lattice distortion essentially located on a single site and which follows instantaneously the vibron. The dressing effect modifies the vibron frequency, reduces the vibron bandwidth and allows for an attractive interaction between vibrons mediated by virtual phonons. Such an interaction is responsible for the formation of bound states and it has been suggested that proteins can support solitons formed by bound states involving a large number of vibrational quanta \cite{kn:ivic2,kn:ivic3}. 

As a consequence, the previous results clearly show that both the intramolecular anharmonicity and the strong vibron-phonon coupling produce a similar effect onto the vibron dynamics and favor the formation of bound states. The present paper is thus devoted to the fundamental question of the interplay between both nonlinear sources. 
To proceed, the Davydov model, modified by introducing the intramolecular anharmonicity of each amide-I vibration, is described within the small polaron approach.
The two-vibron energy spectrum is studied with a special emphasis on the influence of the intramolecular anharmonicity onto the dressing effect. Note that we do not investigate the formation of solitons but focus how attention onto the creation of TVBS, only. Such a procedure is twofold. First, as mentioned previously, TVBS are the first quantum states sensitive to the nonlinearity. Their characterization allows us to understand the interplay between both nonlinear sources and appears as a first step to study the formation of multi-vibron solitons.
Then, recent theoretical calculations have shown that single-vibron solitons do not last long enough to be useful at biological temperatures \cite{kn:cruzeiro1,kn:cruzeiro2}. By contrast, two-vibron solitons are more stable and appear as good candidates for bioenergy transport \cite{kn:cruzeiro3,kn:forner}.
Note that a perfect knowledge of the two-vibron dynamics is also required to interpret some experiments such as time resolved pump-probe spectroscopy \cite{kn:hamm1,kn:hamm2}.
 
The paper is organized as follows. In Sec. II, the Davydov Hamiltonian for a one-dimensional molecular lattice is described. In Sec. III, we first realize a unitary transformation to remove the intramolecular anharmonicity. Then, a modified Lang-Firsov \cite{kn:lang} transformation is applied to renormalize the vibron-phonon interaction and to reach the small polaron point of view. Finally, a mean field procedure is performed to obtain the dressed anharmonic vibron Hamiltonian. In Sec. IV, we summerize the number states method used to solve the two-vibron Schrodinger equation. In Sec. V, a detailed analysis of the two-vibron energy spectrum is performed depending on the values taken by the relevant parameters of the problem. Finally, the results are discussed and interpreted in Sec. VI.
 
\section{The Davydov Hamiltonian}

According to the original Davydov model, the vibron-phonon dynamics in an $\alpha$-helix protein is described by a one-dimensional system formed by $N$ sites periodically distributed along the lattice. Each site is occupied by a peptide group which contains the amide-I vibration. The $n$th amide-I vibration is assumed to behave as an internal high frequency oscillator described by the standard creation and annihilation vibron operators $b^{+}_{n}$ and $b_{n}$. The vibron Hamiltonian is thus written as (using the convention $\hbar$=1)
\begin{eqnarray} 
H_{v}&=& \sum_{n} \omega_{0}b^{+}_{n}b_{n}+\gamma_{3}(b^{+}_{n}
+b_{n})^{3}+\gamma_{4}(b^{+}_{n}+b_{n})^{4} \nonumber \\
& &-J (b^{+}_{n}+b_{n})(b^{+}_{n+1}+b_{n+1}) 
\label{eq:HV}
\end{eqnarray}
where $\omega_{0}$ stands for the internal frequency of the $n$th amide-I mode and where $J$ denotes the lateral hopping constant between nearest neighbor amide-I vibrations. In Eq.(\ref{eq:HV}), $\gamma_{3}$ and $\gamma_{4}$ represent the cubic and quartic anharmonic parameters of each amide-I mode. 

The amide-I vibrations interact with the phonons of the lattice which characterize the collective dynamics of the external motions of the peptide groups. Within the harmonic approximation, each peptide group, with mass $M$, interacts with its nearest neighbor peptide groups via the lateral force constant $W$. Therefore, the phonons correspond to a set of $N$ low frequency acoustic modes, labeled $\{ q \}$, which the Hamiltonian is defined as 
\begin{equation}
H_{p}=\sum_{q}  \Omega_{q}a^{+}_{q}a_{q}
\label{eq:HP}
\end{equation}
where $a^{+}_{q}$ and $a_{q}$ stand for the phonon operators of the $q$th mode with frequency $\Omega_{q}=\Omega_{c}\mid sin(q/2) \mid$, $\Omega_{c}=\sqrt{4W/M}$ denoting the phonon cutoff frequency.

Finally, the vibron-phonon interaction Hamiltonian, which characterizes a random modulation of the internal frequency of each amide-I mode, is expressed as
\begin{equation}
\Delta H_{vp}= \sum_{qn} \frac{1}{2}(\Delta_{n,q}a^{+}_{q}+\Delta_{n,q}^{*}a_{q}) (b^{+}_{n}+b_{n})^{2}
\label{eq:HVP}
\end{equation}
where the coupling constant $\Delta_{n,q}$ is written as
\begin{equation}
\Delta_{n,q}=-i\frac{\Delta_{0}}{\sqrt{N}}\frac{sin(q)}{\sqrt{\mid sin(q/2) \mid}}e^{-iqn}
\label{eq:DELTA}
\end{equation}
In Eq.(\ref{eq:DELTA}), $\Delta_{0}$ is defined in terms of the coupling parameter $\chi$ introduced in the original Davydov model as $\Delta_{0}=\chi (\hbar^{2} M W)^{-1/4}$ ($\hbar$ as be reintroduced to avoid confusion).

The vibron-phonon dynamics is thus described by the full Hamiltonian $H=H_{v}+H_{p}+\Delta H_{vp}$ which slightly differs from the original Davydov model. The Davydov Hamiltonian is recovered from Eqs.(\ref{eq:HV})-(\ref{eq:HVP}) by restricting the full Hamiltonian to vibron-conserving terms, only, and by neglecting the intramolecular anharmonicity of each amide-I vibration.
Although the full Hamiltonian $H$ yields a rather simple model for the protein dynamics, it cannot be solved exactly due to the anharmonic contributions. The following section is thus devoted to its simplification to obtain an effective Hamiltonian describing the vibron dynamics. 
  
\section{Dressed anharmonic vibrons}

\subsection{Renormalization of the intramolecular anharmonicity}

To remove the cubic and quartic intramolecular anharmonicity, the standard procedure described by Kimball et al. \cite{kn:kimball} is used. First, by disregarding the lateral coupling between nearest neighbor amide-I vibrations as well as the vibron-phonon interaction, the procedure consists in performing a unitary transformation $T$ to diagonalize each anharmonic amide-I mode. Then, an approximate Hamiltonian is obtained by applying the transformation on the full Hamiltonian $H$ and be keeping the vibron-conserving terms, only. 

As a result, by using the formula shown in appendix A, the transformed vibron-phonon Hamiltonian $\tilde{H}=THT^{+}$ is expressed as
\begin{eqnarray}
\tilde{H}&=&\sum_{n}(\omega_{0}-2A-B)b^{+}_{n}b_{n}-Ab^{+2}_{n}b_{n}^{2}-Bb^{+}_{n+1}b^{+}_{n}b_{n+1}b_{n}  \nonumber \\
&-&J_{1}[b^{+}_{n}b_{n+1}+h.o.]
-J_{2}[b^{+2}_{n}b_{n+1}^{2}+h.o.] \nonumber \\
&-&J_{3}[b^{+}_{n}(b^{+}_{n}b_{n}+b^{+}_{n+1}b_{n+1})b_{n+1}+h.o.] \nonumber \\
&+&\sum_{qn} (\Delta_{n,q}a^{+}_{q}+\Delta_{n,q}^{*}a_{q}) [(1+2\eta) b^{+}_{n}b_{n}+\eta
b^{+2}_{n}b_{n}^{2}] \nonumber \\
&+&\sum_{q} \Omega_{q}a^{+}_{q}a_{q} 
\label{eq:HTILDE}
\end{eqnarray}
where $A$ denotes the positive anharmonic parameter written as
\begin{equation}
A=30\frac{\gamma_{3}}{\omega_{0}}^{2}-6 \gamma_{4} 
\label{eq:A}
\end{equation}
In Eq.(\ref{eq:HTILDE}), h.o. stands for the corresponding hermitian operator and the different parameters are defined as 
\begin{eqnarray}
B&=&144 J (\frac{\gamma_{3}}{\omega_{0}})^{2} 
; J_{1}=J(1+44(\frac{\gamma_{3}}{\omega_{0}})^{2}-12\frac{\gamma_{4}}{\omega_{0}}) \nonumber \\
J_{2}&=&4 J(\frac{\gamma_{3}}{\omega_{0}})^{2} 
; J_{3}=J(22(\frac{\gamma_{3}}{\omega_{0}})^{2}-12\frac{\gamma_{4}}{\omega_{0}}) \nonumber \\
\eta&=&120(\frac{\gamma_{3}}{\omega_{0}})^{2}-12\frac{\gamma_{4}}{\omega_{0}} 
\label{eq:parameter}
\end{eqnarray}
Note that Eq.(\ref{eq:HTILDE}) was obtained by disregarding the constant term as well as contributions acting onto states involving more than two vibrons.

The unitary transformation allows us to diagonalize each anharmonic amide-I vibration up to the second order with respect to the anharmonic parameters. As a result, the vibron operators describe new vibrational states, called anharmonic vibrons, which are expressed as linear superimpositions of the harmonic states of each amide-I mode. Therefore,  
Eq.(\ref{eq:HTILDE}) clearly shows that the anharmonicity strongly affects the dynamics of these anharmonic vibrons. It first modifies the harmonic part of the Hamiltonian by inducing a redshift of each internal frequency ($\omega_{0} \rightarrow \omega_{0}-2A-B$) and by changing the strength of the lateral interaction ($J \rightarrow J_{1}$). Then, the 
anharmonicity is responsible for the occurrence of coupling terms which break the vibron independence and which directly affect the two-vibron dynamics. The terms 
$-Ab_{n}^{+}b_{n}^{+}b_{n}b_{n}$ and $-Bb_{n}^{+}b_{n+1}^{+}b_{n}b_{n+1}$ yield an attractive interaction between two vibrons and favor their trapping around the same amide-I site and around two nearest neighbor sites, respectively. The contributions proportional to $J_{2}$ characterize hops in the course of which two vibrons, initially located on the same amide-I vibration, realize simultaneously a transition to a nearest neighbor amide-I site. In the same way, the contributions proportional to $J_{3}$ affect single-vibron hops from states formed by two vibrons located onto the same amide-I vibration. Finally, the anharmonicity leads to a 
correction of the vibron-phonon coupling Hamiltonian and Eq.(\ref{eq:HTILDE}) shows that single- and two-vibron states do not experience the same interaction with the phonon bath. 
When compared with the harmonic situation, the coupling between phonons and single-vibron states is enhanced by the anharmonicity, i.e. it is multiplied by the factor $1+2\eta$. Moreover, the terms $\eta b_{n}^{+}b_{n}^{+}b_{n}b_{n}$, which act onto two-vibron states only, show that two vibrons located onto the same amide-I mode interact in a different way with the phonons bath when compared with two vibrons which are far apart. Note that the anharmonic corrections occurring in Eq.(\ref{eq:parameter}) are typically of the order of $A/\omega_{0}$ and represent rather small contributions which have been neglected in our previous works \cite{kn:pouthier1,kn:pouthier2,kn:pouthier3}. Nevertheless, although some changes can effectively be disregarded, such as the modification of the hopping constants, other contributions play a significant role, especially the correction of the vibron-phonon interaction, as it will be shown in the following sections.  

\subsection{Renormalization of the vibron-phonon coupling}

The next step in our procedure is to partially remove the vibron-phonon coupling Hamiltonian by performing a modified Lang-Firsov \cite{kn:lang} transformation. According to the Ivic and co-workers's remarks \cite{kn:ivic2,kn:ivic3}, the vibron-phonon dynamics is dominated by the so-called dressing effect since the non-adiabatic limit is reached. As a result, we consider a "full dressing" and introduce the following unitary transformation
\begin{equation}
U=exp(\sum_{n}Q_{n}[(1+2\eta)b_{n}^{+}b_{n}+\eta b_{n}^{+}b_{n}^{+}b_{n}b_{n}])
\label{eq:U}
\end{equation}
where $Q_{n}$ is defined as
\begin{equation}
Q_{n}=\sum_{q} [\frac{\Delta_{n,q}}{\Omega_{q}} a_{q}^{+}-\frac{\Delta_{n,q}^{*}}{\Omega_{q}} a_{q}]
\label{eq:Q}
\end{equation}
The transformation $U$ (Eq.(\ref{eq:U})) slightly differs from the true Lang-Firsov transformation due to its dependence with respect to the vibron population operators. This dependence originates in the modification of the vibron-phonon interaction mediated by the intramolecular anharmonicity (see Eq.(\ref{eq:HTILDE})) and the original Lang-Firsov transformation is recovered by setting $\eta=0$.
Therefore, by using Eq.(\ref{eq:U}), the transformed Hamiltonian $\hat{H}=U\tilde{H}U^{+}$
is expressed as
\begin{eqnarray}
&&\hat{H}=\sum_{n} \hat{\omega}_{0}b^{+}_{n}b_{n}-\hat{A}b^{+2}_{n}b_{n}^{2}-\hat{B}b^{+}_{n+1}b^{+}_{n}b_{n+1}b_{n}  \nonumber \\
&-&J_{1}[\Theta_{n}^{+}(N_{n}-1)\Theta_{n+1}(N_{n+1})b^{+}_{n}b_{n+1}+h.o.] \nonumber \\
&-&J_{2}[\Theta_{n}^{+2}(N_{n}-\frac{3}{2})\Theta_{n+1}^{2}(N_{n+1}+\frac{1}{2})b^{+2}_{n}b_{n+1}^{2}+h.o.] \nonumber \\
&-&J_{3}[\Theta_{n}^{+}(N_{n}-1)\Theta_{n+1}(N_{n+1})b^{+}_{n}[N_{n}+N_{n+1}]b_{n+1}+h.o.] \nonumber \\
&+&\sum_{q} \Omega_{q}a^{+}_{q}a_{q} 
\label{eq:HHAT}
\end{eqnarray}
where $N_{n}=b_{n}^{+}b_{n}$ denotes the vibron population operator and where the different parameters occurring in Eq.(\ref{eq:HHAT}) are expressed in terms of the small polaron binding energy $E_{B}=2\Delta_{0}^{2}/\Omega_{c}$ as
\begin{eqnarray}
\hat{\omega}_{0}&=&\omega_{0}-2A-B-(1+4\eta)E_{B} \nonumber \\
\hat{A}&=&A+(1+8\eta)E_{B} ;
\hat{B}=B+(1+4\eta)E_{B}
\label{eq:parameter2}
\end{eqnarray}
In Eq.(\ref{eq:HHAT}), $\Theta_{n}(N_{n})$ stands for the dressing operator defined as
\begin{equation}
\Theta_{n}(N_{n})=exp(-Q_{n}[1+2\eta+2\eta N_{n}])
\label{eq:theta}
\end{equation}

In this dressed anharmonic vibron point of view (Eq.(\ref{eq:HHAT})), the vibron-phonon coupling remains through the modulation of the lateral terms by the dressing operators $\Theta_{n}(N_{n})$. Although these operators depend on the phonon coordinates in a highly nonlinear way, the vibron-phonon interaction has been strongly reduced within this transformation. As a result, we can take advantage of such a reduction to express the full Hamiltonian $\hat{H}$ as the sum of three separated contributions as \cite{kn:ivic1,kn:ivic2}
\begin{equation}
\hat{H}=\hat{H}_{eff}+H_{p}+\Delta H 
\end{equation}
where $\hat{H}_{eff}=<\hat{H}>-H_{p}$ denotes the effective Hamiltonian of the dressed anharmonic vibrons and where $\Delta H =\hat{H}-<\hat{H}>$ stands for the remaining part of the vibron-phonon interaction. The symbol $<...>$ represents a thermal average over the phonon degrees of freedom which are assumed to be in thermal equilibrium at temperature $T$.
As a result, the effective dressed anharmonic vibron Hamiltonian is written as
\begin{eqnarray}
&&\hat{H}_{eff}=\sum_{n} \hat{\omega}_{0}b^{+}_{n}b_{n}-\hat{A}b^{+2}_{n}b_{n}^{2}-\hat{B}b^{+}_{n+1}b^{+}_{n}b_{n+1}b_{n}  \nonumber \\
&&-J_{1}[\Phi(N_{n}+N_{n+1})b^{+}_{n}b_{n+1}+h.o.] \nonumber \\
&&-J_{2}[\Phi(N_{n}+N_{n+1})^{4}b^{+2}_{n}b_{n+1}^{2}+h.o.] \nonumber \\
&&-J_{3}[\Phi(N_{n}+N_{n+1})b^{+}_{n}[N_{n}+N_{n+1}]b_{n+1}+h.o.] 
\label{eq:HHATEFF}
\end{eqnarray}
where $\Phi(X)=exp(-S(T)[1+2\eta+2\eta X])$ and where $S(T)$ is the coupling constant 
introduced by Ivic and co-workers as ($k_{B}$ denotes the Boltzmann constant)
\begin{equation}
S(T)=\frac{4E_{B}}{N\Omega_{c}} \sum_{q} \mid sin(\frac{q}{2}) \mid cos(\frac{q}{2})^{2} coth(\frac{\Omega_{q}}{2k_{B}T})
\label{eq:S}
\end{equation}

The Hamiltonian $\hat{H}_{eff}$ (Eq.(\ref{eq:HHATEFF})) describes the dynamics of the anharmonic vibrons dressed by a virtual cloud of phonons, i.e. anharmonic small polarons.
It takes into account on the anharmonicity up to the second order and allows for a renormalization of the main part of the vibron-phonon coupling within the non-adiabatic limit. The interaction Hamiltonian $\Delta H$, which characterizes the coupling between these dressed anharmonic vibrons and the remaining phonons, is assumed to be small in order to be treated using perturbation theory. Such a contribution, responsible for phase relaxation, will be studied in a forthcoming paper. 

Eq.(\ref{eq:HHATEFF}) clearly shows the interplay between the intramolecular anharmonicity and the strong vibron-phonon coupling. The Hamiltonian $\hat{H}_{eff}$ exhibits, basically, the same contributions as the vibron part of the anharmonic Hamiltonian $\tilde{H}$. The main difference is that the parameters occurring in $\hat{H}_{eff}$ are renormalized due to the dressing effect which modifies the dynamics according to two main ways. 
First, it yields additional contributions to the anharmonic parameters $A$ and $B$ as well as to the internal frequency $\omega_{0}$ (Eq.(\ref{eq:parameter2})). Then, the dressing effect modifies the different lateral contributions via the dressing function $\Phi(X)$ which reduces the hopping constants. However, in a marker contrast with the harmonic situation, the intramolecular anharmonicity enhances the role played by both the small polaron binding energy and the coupling constant $S(T)$. In addition, the dressing effect depends on the vibron population. As shown in the next section, such a dependence discriminates transitions involving two vibrons located onto the same amide-I vibration from transitions involving two vibrons which are far apart.

\section{Two-vibron states}

Since the effective Hamiltonian $\hat{H}_{eff}$ (Eq.(\ref{eq:HHATEFF})) conserves the number of dressed anharmonic vibrons, its eigenstates can be characterized by using the number states method \cite{kn:scott,kn:pouthier1,kn:pouthier2,kn:pouthier3}. Within this method, the two-vibron wave function is expanded as  
\begin{equation}
\mid \Psi \rangle = \sum_{n_{1},n_{2}\geq n_{1}} \Psi(n_{1},n_{2}) 
\mid n_{1},n_{2} ) 
\label{eq:basis}
\end{equation}
where $\{ \mid n_{1},n_{2} ) \}$ denotes a local basis set normalized and symmetrized according to the restricting $n_{2} \geq n_{1}$ due to the indistinguishable nature of the vibrons \cite{kn:pouthier1,kn:pouthier2,kn:pouthier3}. A particular vector $\mid n_{1},n_{2} )$ characterizes two vibrons located onto the sites $n_{1}$ and $n_{2}$, respectively.
The expression of the corresponding time independent Schrodinger equation, 
$\hat{H}_{eff}\mid \Psi \rangle =\omega \mid \Psi \rangle$, depends on the nature of the basis vectors involved in. Indeed, from Eq.(\ref{eq:HHATEFF}), there are three different situations either describing two vibrons located onto sites which are far apart, two vibrons located onto nearest neighbor sites and two vibrons located onto the same site. 

In the first situation, the Schrodinger equation is expressed as 
\begin{eqnarray}
&-&J_{1}\Phi(1)[\Psi(n_{1},n_{2}+1)+\Psi(n_{1},n_{2}-1)] \nonumber \\
&-&J_{1}\Phi(1)[\Psi(n_{1}+1,n_{2})+\Psi(n_{1}-1,n_{2})] \nonumber \\
&+&2\hat{\omega}_{0}\Psi(n_{1},n_{2})=\omega\Psi(n_{1},n_{2})
\label{eq:schrod1}
\end{eqnarray}
Eq.(\ref{eq:schrod1}) shows that the two vibrons move in an independent way according to an effective hopping constant $J_{1}\Phi(1)=J_{1}exp[-(1+4\eta)S(T)]$ which slightly differs from the hopping constant involved in the harmonic approximation. Indeed, as discussed in the previous section, the intramolecular anharmonicity modifies the harmonic hopping constant ($J \rightarrow J_{1}$) and enhances the dressing effect ($S(T)\rightarrow (1+4\eta)S(T)$). Therefore, although the anharmonic parameter $\eta$ is rather small, it acts under the exponential and then reduces the effective hopping constant with respect to the harmonic situation. 

When two vibrons are located onto nearest neighbor amide-I vibrations, the Schrodinger equation is expressed as
\begin{eqnarray}
&-&J_{1}\Phi(1)[\Psi(n_{1},n_{1}+2)+\Psi(n_{1}-1,n_{1}+1)] \nonumber \\
&-&\sqrt{2}(J_{1}+J_{3})\Phi(2)[\Psi(n_{1},n_{1})+\Psi(n_{1}+1,n_{1}+1)] \nonumber \\
&+&(2\hat{\omega}_{0}-\hat{B})\Psi(n_{1},n_{1}+1)=\omega\Psi(n_{1},n_{1}+1)
\label{eq:schrod2}
\end{eqnarray}
and, when two vibrons are located onto the same amide-I mode, the Schrodinger equation is defined as
\begin{eqnarray}
&-&2J_{2}\Phi(2)^{4}[\Psi(n_{1}+1,n_{1}+1)+\Psi(n_{1}-1,n_{1}-1)] \nonumber \\
&-&\sqrt{2}(J_{1}+J_{3})\Phi(2)[\Psi(n_{1},n_{1}+1)+\Psi(n_{1}-1,n_{1})] \nonumber \\
&+&(2\hat{\omega}_{0}-2\hat{A})\Psi(n_{1},n_{1})=\omega\Psi(n_{1},n_{1})
\label{eq:schrod3}
\end{eqnarray}
As shown in Eqs.(\ref{eq:schrod2})-(\ref{eq:schrod3}), both the intramolecular anharmonicity and the dressing effect strongly modify the dynamics involving vibrons located onto neighboring sites. First, they contribute to a redshift of the energies of the corresponding states. Then, they affect the hopping processes by favoring simultaneous motions of two vibrons and by changing the dressing effect which is characterized by the effective correction $\Phi(2)=exp[-(1+6\eta)S(T)]$. This latter feature originates in the dependence of the dressing operators (Eq.(\ref{eq:theta})) on the vibron population. Therefore, since $\Phi(2)$ is lesser than $\Phi(1)$, it is clear that the trapping process experienced by two vibrons located onto the same site is more efficient than the dressing effect which modify the single-vibron dynamics. 

Eqs.(\ref{eq:schrod1})-(\ref{eq:schrod3}) clearly indicate how the physics of the two-vibron states is related to the dynamics of a single fictitious particle moving quantum mechanically on the two-dimensional lattice displayed in Fig. 1a \cite{kn:pouthier1,kn:pouthier2,kn:pouthier3}. 
Within this equivalence, the two-vibron wave function $\Psi(n_{1},n_{2})$ can be viewed as the wave function of the fictitious particle. Its dynamics is described by a tight-binding Hamiltonian characterized by the self-energy $2\hat{\omega}_{0}$ located on each site and a hopping matrix $J_{1}\Phi(1)$ which couples nearest neighbor sites. However, the 2D lattice exhibits two rows of defects (see Fig. 1a) which yield
a redshift of the self-energy of the corresponding sites as well as for a modification of the hopping matrix elements connecting the defect sites. Therefore, such defects allow us to discriminate between two different eigenstates. The eigenstates of the core of the lattice  correspond to plane waves slightly perturbed by the defects. By contrast, the presence of the defects leads to the occurrence of states which are localized in the vicinity of the two defect rows. In terms of the two-vibron dynamics, the previous equivalence yields two different kinds of eigenstates. Indeed, the delocalization of the fictitious particle is associated to a free motion of the two vibrons, i.e. two-vibron free states (TVFS), whereas its localization is connected to the occurrence of TVBS. Note that in marked contrast with the standard Hubbard Hamiltonian for bosons \cite{kn:kimball,kn:scott,kn:pouthier2}, the presence of two defect rows in our equivalent lattice shows that the system can support two kinds of bound states corresponding to the trapping of the two vibrons either onto the same amide-I vibration or onto nearest neighbor amide-I sites, respectively. 

The Schrodinger equation Eqs.(\ref{eq:schrod1})-(\ref{eq:schrod3}) can be expressed in an improved way by taking advantage of the lattice periodicity.
Indeed, the two-vibron wave function is invariant with respect to a translation 
along the lattice and can be expanded as a Bloch wave as
\begin{equation}
\Psi(n_{1},n_{2}=n_{1}+m)= \frac{1}{\sqrt{N}} \sum_{n_{1}} e^{ik(n_{1}+m/2)}\Psi_{k}(m)
\label{eq:bloch}
\end{equation}
In Eq.(\ref{eq:bloch}), the total momentum $k$, which belongs to the first Brillouin zone of the molecular lattice, is associated to the motion of the center of mass of the two vibrons whereas the resulting wave function $\Psi_{k}(m)$ refers to the degree of freedom $m$ which characterizes the distance between the two vibrons. 
Since the momentum $k$ is a good quantum number, the Hamiltonian $\hat{H}_{eff}$ appears as block diagonal and the Schrodinger equation can be solved for each $k$ value. Therefore, when two vibrons are located onto sites which are far apart ($m > 1$), Eq.(\ref{eq:schrod1}) becomes
\begin{equation}
(2\hat{\omega}_{0}-\omega)\Psi_{k}(m) -\Gamma_{k}[\Psi_{k}(m+1)+\Psi_{k}(m-1)] = 0
\label{eq:schrod11}
\end{equation}
where $\Gamma_{k}=2J_{1}\Phi(1)cos(k/2)$. In the same way, when two vibrons are located onto nearest neighbor sites ($m=1$), Eq.(\ref{eq:schrod2}) is rewritten as 
\begin{equation}
(2\hat{\omega}_{0}-\hat{B}-\omega)\Psi_{k}(1)-\Gamma_{k}\Psi_{k}(2)
-\sqrt{2}\gamma_{k}\Psi_{k}(0)=0
\label{eq:schrod22}
\end{equation}
where $\gamma_{k}=2(J_{1}+J_{3})\Phi(2) cos(k/2)$. Then, when two vibrons are located onto the same site ($m=0$), Eq.(\ref{eq:schrod3}) is expressed as 
\begin{equation}
(2\hat{\omega}_{0}-2\hat{A}_{k}-\omega)\Psi_{k}(0) - \sqrt{2}\gamma_{k}\Psi_{k}(1)=0
\label{eq:schrod33}
\end{equation}
where $\hat{A}_{k}=\hat{A}+2J_{2} \Phi(2)^{4} cos(k)$.

Finally, for each wave vector $k$, the two-vibron dynamics reduces to a one-dimensional tight-binding problem on a semi-infinite lattice which exhibits two defects (see Fig. 1b). The Schrodinger equation Eqs.(\ref{eq:schrod11})-(\ref{eq:schrod33}) can thus be solved easily to obtain the two-vibron eigenstates and to determine the two-vibron energy spectrum. This procedure is illustrated in the following section. 

\section{Numerical results}

In this section, the previous formalism is applied to characterize the two-vibron energy spectrum of an $\alpha$-helix protein. However, the present theory involves a set of parameters which has first to be discussed. 

From the literature, the harmonic dynamics of vibrons in $\alpha$-helices is relatively well described (see for instance Refs. \cite{kn:scott,kn:ivic2}). The quantum energy for an amide-I vibration is about $\omega_{0}=1665$ cm$^{-1}$ and a well admitted value for the hopping constant is $J=7.8$ cm$^{-1}$. By contrast, the phonon dynamics as well as the vibron-phonon coupling parameter are only partially known. The mass $M$ of a peptide group ranges between 1.17 $10^{-25}$ kg and 1.91 $10^{-25}$ kg, whereas the phonon force constant $W$ is expected to be about 13-19.5 Nm$^{-1}$. As a result, the phonon cutoff frequency varies from 87 to 137 cm$^{-1}$. In the same way, a typical range for the vibron-phonon coupling term is $\chi=35-62$ pN. Therefore, the vibron-phonon coupling parameter $\Delta_{0}$ ranges between 12 and 29 cm$^{-1}$ and the small polaron binding energy $E_{B}$ extends from 3 to 15 cm$^{-1}$. 

In a marked contrast with the previous parameters, little is known about the intramolecular anharmonicity of each amide-I vibration. Therefore, to determine the range of the anharmonic parameters, we assume that the amide-I vibration is equivalent to the C=O stretching mode of a single molecule, but with a different reduced mass $\mu$. By describing both vibrations according to a Morse potential, it is straightforward to show that $\omega_{0}$ scales as $1/\sqrt{\mu}$ whereas the anharmonicity $A$ scales as $1/\mu$. By comparing the harmonic frequency of both the amide-I vibration and the C=O stretching mode ($\omega_{0}=2170$ cm$^{-1}$ \cite{kn:herzberg}), the reduced mass of the amide-I mode is about 1.6-1.7 times the reduced mass of the C=O molecule. As a consequence, since the anharmonic parameter for the C=O stretching mode is $A=13.3$ cm$^{-1}$ \cite{kn:herzberg}, the anharmonic constant of the amide-I vibration is about $A=8.0$ cm$^{-1}$. Note that this latter value represents an order of magnitude of the anharmonic parameter since, strictly speaking, the anharmonicity of the amide-I vibration depends on the details of the corresponding intramolecular potential.    
In addition, our calculations establish that both the cubic and quartic anharmonic parameters can be expressed approximately in terms of $A$ by using the relation $15\gamma_{3}^{2}/\omega_{0} \approx 6 \gamma_{4}\approx A$. 

As a result, in the following of the text, the intramolecular anharmonicity will be described by a single parameter, namely the anharmonic constant $A$. The small polaron binding energy $E_{B}$ is taken as a parameter whereas the phonon cutoff frequency $\Omega_{c}$ is fixed to 100 cm$^{-1}$. The temperature is set to the biological temperature, i.e. $T=310$ K, and the hopping constant is set to $J=7.8$ cm$^{-1}$.

The influence of the intramolecular anharmonicity on the two-vibron energy spectrum is shown in Figs. 2, 3 and 4 for three typical situations. In each figure, the spectrum is centered onto the corrected frequency $2\hat{\omega}_{0}$ (Eq.(\ref{eq:parameter2})) and corresponds to the two-vibron dispersion curves drawn in the first Brillouin zone of the lattice, i.e. $-\pi<k<\pi$. 

When $E_{B}=8$ cm$^{-1}$ and $A=0$ (Fig. 2a), the two-vibron energy spectrum is formed by an energy continuum associated to the TVFS. This continuum is symmetrically located around $2\hat{\omega}_{0}$ with a bandwidth equal to 31.21 cm$^{-1}$. Below the continuum, the energy spectrum exhibits two bands connected to two different bound states. The low frequency band, located below the TVFS over the entire Brillouin zone, refers to bound states denoted TVBS-I. The bottom of this band is located at 24.74 cm$^{-1}$ below the center of the continuum and its bandwidth is about 8.7 cm$^{-1}$. The second band, which refers to bound states denoted TVBS-II, is located below the continuum at the end of the Brillouin zone, only. The band disappears inside the continuum when $\mid k \mid < k_{c}=2.06$. When the anharmonic parameter is set to $A=10$ cm$^{-1}$ (Fig. 2b), the TVFS bandwidth is reduced to 28.40 cm$^{-1}$. Note that the center of the continuum is redshifted due to the dependence of $2\hat{\omega}_{0}$ (Eq.(\ref{eq:parameter2})) on the anharmonic parameter $A$ (not drawn in the figure). Then, the intramolecular anharmonicity is responsible for a strong redshift of the TVBS-I band as well as for a decrease of the corresponding bandwidth. The bottom of the band is located at 43.39 cm$^{-1}$ below the center of the continuum and the bandwidth is equal to 2.78 cm$^{-1}$. Finally, although the anharmonicity does not significantly change the shape of the TVBS-II band, it leads to a decrease of the critical wave vector for which its disappearance arises since $k_{c}=0.70$. 

When $E_{B}$ is set to 10 cm$^{-1}$, the two-vibron energy spectrum is shown in Figs. 3a and 3b. In the harmonic case, i.e. $A=0$ (Fig. 3a), the energy spectrum exhibits the same features as in Fig. 2a. Nevertheless, the TVFS bandwidth is smaller and equal to 26.24 cm$^{-1}$. The width of the TVBS-I band has been reduced to 6.04 cm$^{-1}$ and the bottom of the band is located at 26.04 cm$^{-1}$ below the center of the continuum. In addition, the critical wave vector connected to the disappearance of the TVBS-II has been reduced to $k_{c}=1.70$. When the anharmonic parameter is set to $A=10$ cm$^{-1}$, we observe the same behavior as in Fig. 2b. Indeed, both the TVFS and the TVBS-I bandwidths decrease to reach 23.30 cm$^{-1}$ and 1.70 cm$^{-1}$, respectively. In addition, the bottom of the TVBS-I band is strongly redshifted and is located at 47.47 cm$^{-1}$ below the center of the continuum. However, a new process arises since the intramolecular anharmonicity yields the occurrence of the TVBS-II band over the entire Brillouin zone. The corresponding bandwidth is equal to 1.53 cm$^{-1}$ and the bottom of the band is located at 13.42 cm$^{-1}$ below the center of the continuum.  

The situation corresponding to a strong small polaron binding energy is finally illustrated in Figs. 4a and 4b for $E_{B}$=14 cm$^{-1}$. The main difference when compared with the two previous cases is that the dressing effect is strong enough to induce the occurrence of the TVBS-II band over the entire Brillouin zone, even when $A=0$ (Fig. 4a). As previously, the intramolecular anharmonicity reduces both the TVFS continuum and the TVBS-I bandwidth and yields a strong redshift of the TVBS-I band. Moreover, it increases the frequency difference between the TVBS-II band and the center of the continuum and reduces the TVBS-II bandwidth. 

The difference between the two bound states is illustrated in Fig. 5. Figs. 5a and 5b show the TVBS wave functions with zero wave vector and which the energy spectrum is displayed in Figs. 3a and 3b, respectively ($E_{B}=10$ cm$^{-1}$). By contrast, Figs. 5c and 5d refer to the TVBS which the energy spectrum is shown in Figs. 4a and 4b, respectively ($E_{B}=14$ cm$^{-1}$). The TVBS-I wave function is described by open circles whereas the TVBS-II wave function is represented by open squares. When $A=0$ and $E_{B}=10$ cm$^{-1}$ (Fig. 5a), the TVBS-I wave function is maximum when $m=0$ and decreases as the separating distance $m$ between the two vibrons increases. However, it does not exhibit a true exponential decrease versus $m$. When the anharmonicity is set to $A=10$ cm$^{-1}$ (Fig. 5b), the lattice supports two bound states with zero wave vector (see Fig. 3b). As previously, the TVBS-I wave function is maximum when $m=0$ but the extension of the wave function has been reduced when compared with the harmonic case, i.e. it decreases exponentially versus $m$. By contrast, the TVBS-II wave function is maximum when $m=1$ and exhibits an exponential tail as $m$ increases. When $E_{B}=14$ cm$^{-1}$, the lattice supports two bound states with zero wave vector whatever the anharmonicity (Figs. 4a and 4b). When $A=0$ (Fig. 5c), the TVBS-I wave function, which is maximum when $m=0$, shows a significant value when $m=1$ and does not decrease according to a true exponential. The TVBS-II wave function is maximum when $m=1$ and exhibits a rather important weight when $m=0$. By contrast, when the anharmonicity is set to $A=10$ cm$^{-1}$ (Fig. 5d), the TVBS-I and the TVBS-II wave functions are clearly localized onto $m=0$ and $m=1$, respectively, both wave functions exhibiting an exponential tail versus $m$. 

As mentioned previously, the intramolecular anharmonicity is responsible for a slight decrease of the TVFS bandwidth. This feature is illustrated in Fig. 6 where the behavior of the bandwidth is shown as a function of $A$ for different values of $E_{B}$. The figure exhibits two distinct regimes. When $E_{B}=0$, the TVFS bandwidth slightly increases in a linear way as the anharmonicity increases. The bandwidth, equal to 62.40 cm$^{-1}$ when $A=0$, is blueshifted of about 0.36 cm$^{-1}$ when $A=10$ cm$^{-1}$. In a marked contrast, for finite values of $E_{B}$, the TVFS bandwidth is strongly reduced and decreases as $A$ increases. When $E_{B}=9$ cm$^{-1}$ and $A=0$, the bandwidth is equal to 28.6 cm$^{-1}$ whereas it reaches the value 25.7 cm$^{-1}$ when $A=10$ cm$^{-1}$, i.e. a variation of about 2.9 cm$^{-1}$. 

In Figs. 7a and 7b, the influence of the anharmonicity on the TVBS-I is illustrated for different values of $E_{B}$. As shown in Fig. 7a, the TVBS-I binding energy $E_{I}$, defined as the gap between the TVBS-I with zero wave vector and the bottom of the TVFS continuum, decreases as the anharmonicity increases. When $E_{B}=0$, $E_{I}$ decreases from zero according to a $A^{2}$ power law for small $A$ values and reaches a quasi-linear decrease for a stronger anharmonicity. For finite values of $E_{B}$, $E_{I}$ reaches more rapidly the linear regime as increasing $A$ and the slope of the decrease slightly increases with $E_{B}$. Note that $E_{I}$ decreases with the small polaron binding energy. In Fig. 7b, the behavior of the TVBS-I bandwidth as function of $A$ is illustrated for different values of $E_{B}$. In a general way, the TVBS-I bandwidth decreases as the anharmonicity increases. However, for a small anharmonicity, the TVBS-I bandwidth exhibits a rather fast decrease whereas it decreases more slowly for a stronger anharmonicity. When $E_{B}=9$ cm$^{-1}$, the TVBS-I bandwidth is equal to 7.28 cm$^{-1}$ when $A=0$ and reaches  2.67 cm$^{-1}$ when $A=8$ cm$^{-1}$, i.e. almost 0.3 times smaller.

Finally, Fig. 8 displays the behavior of the critical wave vector $k_{c}$ for which the TVBS-II band disappears. In a general way, $k_{c}$ decreases as A increases and vanishes for a critical value of the anharmonic parameter $A=A_{c}$. In others words, when $A>A_{c}$, the TVBS-II band is located below the continuum. By contrast, when $A<A_{c}$, the TVBS-II band appears below the continuum at the end of the Brillouin zone only, i.e. when $\mid k \mid > k_{c}$. Note that $A_{c}$ decreases as $E_{B}$ increases. Moreover, the different curves clearly exhibit a critical behavior since $k_{c}$ scales as $k_{c} \approx (A-A_{c})^{1/2}$ when it approaches zero. Note that, as shown in the inset of Fig. 8, $k_{c}$ shows a similar behavior with respect to the temperature. Indeed, for $A=8$ cm$^{-1}$ and $E_{B}=10$ cm$^{-1}$, the critical wave vector decreases as increasing the temperature and vanishes when $T=T_{c}=236$ K. Therefore, when $T>T_{c}$ , the TVBS-II band is located below the continuum over the entire Brillouin zone. Note that $k_{c}$ shows the same critical behavior when it approaches zero and scales as $k_{c} \approx (T-T_{c})^{1/2}$.
 
\section{Discussion}

To interpret and discuss the previous numerical results, let us first focus our attention onto the influence of the anharmonicity on the TVFS continuum. As shown in Figs. 2, 3, 4 and 6, the intramolecular anharmonicity is responsible for a redshift of the TVFS bandwidth, typically of about a few wave numbers. Note that, when $E_{B}=0$, i.e. when there is no coupling with the phonon bath, the anharmonicity leads to a very small blueshift of the continuum bandwidth. These features originate in the modification of the hopping constant $J$ due to both the anharmonicity and the dressing effect. Indeed, as discussed in Sec. IV, a TVFS corresponds to an independent motion of the two vibrons according to an effective hopping constant $J_{1}\Phi(1)=J_{1}exp[-(1+4\eta)S(T)]$ which differs from the constant involved in the harmonic approximation. First, the intramolecular anharmonicity slightly increases the harmonic hopping constant (see Eq.(\ref{eq:parameter})) yielding a blueshift of the TVFS continuum when $E_{B}=0$. Then, the anharmonicity reinforces the role played by the coupling constant $S(T)$ which characterizes the dressing effect ($S(T)\rightarrow (1+4\eta)S(T)$). Therefore, it enhances the dressing effect and favors a decrease of the effective hopping constant. Although both contributions act in an opposite way, the enhancement of the dressing effect appears to be more efficient for non vanishing values of the small polaron binding energy. In other words, anharmonic vibrons are more sensitive to the dressing effect than harmonic vibrons. For instance, when $T=310$ K, $E_{B}=14$ cm$^{-1}$ and $J=7.8$ cm$^{-1}$, the effective hopping constant $J_{1}\Phi(1)$ is equal to 2.32 cm$^{-1}$ within the harmonic case whereas it is redshifted to 1.96 cm$^{-1}$ when $A=10$ cm$^{-1}$, i.e. a reduction of about 15\%. 

The main result of the present study concerns the modification of the bound states due to the interplay between the intramolecular anharmonicity and the strong vibron-phonon coupling. Indeed, for the harmonic lattice, the dressing effect is responsible for the occurrence of two different bound states, i.e. TVBS-I and TVBS-II. The TVBS-I are located below the TVFS continuum over the entire Brillouin zone. By contrast, for TVBS-II, two situations occur depending on the strength of the small polaron binding energy. For small values of $E_{B}$, the band disappears inside the continuum when the wave vector $\mid k \mid$ is lower than a critical wave vector $k_{c}$ whereas, for strong values of $E_{B}$, the band is located below the continuum over the entire Brillouin zone. As when increasing the anharmonicity, the TVBS-I band is redshifted and its bandwidth is strongly reduced. In the same way, the anharmonicity modifies the nature of the TVBS-II band. If the band is resonant with the continuum, the anharmonicity yields a decrease of the critical wave vector. Therefore, if it is strong enough, the anharmonicity allows for the TVBS-II band to get out of the continuum over the entire Brillouin zone. When  the TVBS-II band is not resonant with the continuum, the anharmonicity induces a redshifted of the band as well as a decrease of its bandwidth. All the previous features are accompanied by a modification of the wave functions of the bound states.

To understand these features, we can take advantage of the equivalence between the two-vibron dynamics and the tight-binding problem on the one-dimensional lattice displayed in Fig. 1b. Within this equivalence, bound states in the real molecular lattice are described in terms of localized states in the equivalent lattice, the localization occurring due to the presence of the two defects. For a given $k$ value, the previous results clearly show that the system supports one or two bound states, depending on the values taken by the relevant parameters of the problem (anharmonicity, small polaron binding energy, temperature ...). In other words, if these parameters are allowed to vary, a localization transition arises in the equivalent lattice. Such a transition discriminates between two "phases". The first phase corresponds to the presence of a single localized state (i.e. a single bound state in the real lattice) whereas the second phase is connected to the occurrence of two localized states (i.e. two bound states). A localized state is characterized by its localization length $\xi $
which refers to the length of the bond between two vibrons in a bound state. The disappearance of a localized state is accompanied by a divergence of the localization length since it now refers to two independent vibrons. 

Such a process is similar to a critical transition and consequently, we can use the tools of the Renormalization Group (RG) theory to understand the transition and predict the occurrence of localized states (i.e. of bound states). This can be achieved by performing a decimation of the Schrodinger equation as detailed in Refs. \cite{kn:pouthier4,kn:pouthier5} and summarized in appendix B. The decimation consists in eliminating from the initial equivalent lattice one site over two to arrive at a scaled lattice with twice the lattice spacing. If the initial lattice is at a critical point, i.e. if the localization transition takes place, the localization length $\xi $
is infinite and the dynamics is in a self-similar situation. 
As a result, no change in the critical
parameters accompanies the length scaling and the scaled lattice remains at
a critical point. The decimation procedure can be applied recursively until
the lattice parameter has been increased up to infinity. 
By operating in such a way, we decrease drastically the number of sites in
the equivalent lattice, and obtain a scaled lattice which reduces to the two side sites $m=0$ and $m=1$, only (see Fig. 1b). It is thus possible to study exactly the Schrodinger equation of such a critical system and then to characterize the critical properties of the
initial lattice. By following this procedure (see appendix B), the condition for the occurrence of localized states in the equivalent lattice, i.e. for the occurrence of bound states in the real lattice, is defined as 
\begin{equation}
(\hat{A}_{k}-\Gamma_{k})(\hat{B}-\Gamma_{k})=\gamma_{k}^{2}
\label{eq:phase}
\end{equation}
where the parameters occurring in Eq.(\ref{eq:phase}) are defined in Sec. IV.

As shown in Fig. 9, Eq.(\ref{eq:phase}) defines the critical curve in the space of the parameters which separate the phase I, with a single bound state, from the phase II, with two bound states. In phase I, the nature of the bound state depends on the relative values of the parameters and three main situations occur. When $\hat{A}_{k} >> \hat{B}$, the localization arises around the first site $m=0$. In other words, TVBS-I characterizes the trapping of two vibrons around the same amide-I vibration. By contrast, when $\hat{A}_{k} << \hat{B}$, the localization occurs on the second site $m=1$ and the two vibrons are trapped onto nearest neighbor amide-I modes. In the intermediate case, when 
$\hat{A}_{k} \approx \hat{B}$, the localized state is a superimposition of states localized on sites $m=0$ and $m=1$. As a result, TVBS-I characterizes a superimposition of states trapped  over both the same amide-I mode and two nearest neighbor amide-I modes. In phase II, the lattice supports two bound states. When $\hat{A}_{k} >> \hat{B}$, TVBS-I corresponds to the trapping of the two vibrons onto the same amide-I mode whereas TVBS-II characterizes the trapping of the two vibrons onto nearest neighbor amide-I modes. As shown in Fig. 9, the revert situation takes place when $\hat{A}_{k} << \hat{B}$. Finally, when $\hat{A}_{k} < \hat{B}< 2\hat{A}_{k} $, both bound states appear as combinations of states involving the trapping of the two vibrons onto the same amide-I mode and onto nearest neighbor amide-I modes.

In the harmonic $\alpha$-helix protein, i.e. $A=0$, Eq.(\ref{eq:parameter2}) yields $\hat{A}_{k}=\hat{B}=E_{B}$. Therefore, the critical curve Eq.(\ref{eq:phase}) shows that when $E_{B}<4J\Phi(0)cos(k/2)$, the lattice exhibits a single bound state which mixes the trapping onto the same amide-I mode and onto two nearest neighbor amide-I modes (see Fig. 5a). By contrast, when $E_{B}>4J\Phi(0)cos(k/2)$, the lattice supports two bound states, both being a superimposition of states involving the trapping onto the same and onto two nearest neighbor amide-I modes (see Fig. 5c). Note that the condition for the occurrence of two bound states over the entire Brillouin zone is $E_{B}>4J\Phi(0)$. As when increasing the anharmonicity $A$, Eq.(\ref{eq:parameter2}) shows that $\hat{A}_{k}>\hat{B}$. For instance, when $T=310$ K, $E_{B}=14$ cm$^{-1}$ and $A=10$ cm$^{-1}$, $\hat{A}$ is equal to 28.29 cm$^{-1}$ whereas $\hat{B}$ is equal to 16.54 cm$^{-1}$. Therefore, the anharmonicity decreases the hybridization between the two kinds of trapping. As a result, TVBS-I corresponds essentially to the trapping onto the same amide-I mode whereas TVBS-II, when it is present, characterizes the trapping onto two nearest neighbor amide-I modes (see Figs. 5b and 5d). 
Note that the hybridization between the two trapping processes is also reduced 
when the hopping constants $\Gamma_{k}$ and $\gamma_{k}$ decrease, i.e. when the different sites $m$ become uncorrelated. This feature arises when the TVBS wave vector approaches $\pi$ as well as when the temperature increases due to the dressing effect. 

Although the phase diagram displayed in Fig. 9 allows for a complete understanding of the nature of the bound states, it involves parameters which are not independent. For instance, both $\hat{A}_{k}$ and $\hat{B}$ depend on $A$ and $E_{B}$. However, from Eq.(\ref{eq:phase}), we can define a phase diagram in the $(A,E_{B})$ parameter space. Such diagram is illustrated in Fig. 10 for a zero wave vector. The temperature is fixed to $T=310$ K whereas three different values for the hopping constant $J$ have been considered. The curve discriminates between a phase with a single bound state and a phase with two bound states, both states being located below the continuum over the entire Brillouin zone since $k=0$. The figure clearly shows that the anharmonicity favors the occurrence of two bound states by decreasing the value of the required small polaron binding energy. For instance, when $J=7.8$ cm$^{-1}$ and $A=0$, $E_{B}$ must exceed 11.5 cm$^{-1}$ to allow for the occurrence of the TVBS-II. By contrast, this value is reduced to 8.7 cm$^{-1}$ when $A=8$ cm$^{-1}$. Note that the curve is pushed down as decreasing the hopping constant as well as increasing the temperature due to the dressing effect. 

From a physical point of view, the previous phase diagrams can be easily understood in terms of localization in the equivalent lattice. Indeed, since the equivalent lattice exhibits two defects onto the sites $m=0$ and $m=1$, it can supports two states localized onto the sites $m=0$ and $m=1$. However, due to the hopping processes, these two localized states overlapp and interact to generate new localized states which appear as superimpositions of the two previous states. Such a mechanism is accompanied by a splitting of the frequency of the new localized states which depends on both the strength of the coupling between the original states as well as on their frequency. As a result, we obtain a high frequency localized state (TVBS-II) and a low frequency localized state (TVBS-I). If the splitting is strong enough, the high frequency localized state is pushed back into the continuum and a single localized state remains. By contrast, if the splitting is rather weak, both localized states can be located below the continuum. As shown in Eqs.(\ref{eq:schrod11})-(\ref{eq:schrod22}), the intramolecular anharmonicity increases the energy difference between the two original localized states. In addition, since it enhances the dressing effect, it reduces the coupling between these two states. As a result, the anharmonicity favors a weak hybridization between the two original localized states. The main consequence is that TVBS-I and TVBS-II are essentially localized around $m=0$ and $m=1$, respectively. 

The present paper has clearly established the interplay between the intramolecular anharmonicity and the strong vibron-phonon coupling. We have shown that the anharmonicity modifies the vibron-phonon interaction which results in an enhancement of the dressing effect. Therefore, anharmonic vibrons are more sensitive to the dressing than harmonic vibrons and are 
characterized by smaller effective hopping constants. In addition, we have shown that both nonlinear sources break the vibron independence and favor the occurrence of two kinds of bound states which the properties strongly depend on the interplay between the anharmonicity and the dressing effect. Such a dependence was summarized in a phase diagram which characterizes the number as well as the nature of the bound states as a function of the values taken by the relevant parameters of the problem. In the harmonic situation, the two bound states appear as combinations of states involving the trapping of the two vibrons onto the same amide-I mode and onto nearest neighbor amide-I modes. By contrast, the intramolecular anharmonicity reduces the hybridization between these two kinds of trapping so that the low frequency bound state refers to the trapping of the two vibrons onto the same amide-I mode whereas the high frequency bound state characterizes the trapping onto nearest neighbor amide-I vibrations. In addition, the anharmonicity strongly reduces the dispersion of the bound states and thus enhances their breather-like behavior \cite{kn:pouthier3}.

To conclude, let us mention that forthcoming works will be devoted to the fundamental question of the TVBS lifetime due to the coupling $\Delta H$ (Eq. (17)) with the remaining phonons. Such a problem was studied in a recent paper \cite{kn:pouthier2} in which the decay of TVBS into either bound or free states was described by considering a weak vibron-phonon coupling. However, in the present context, both the anharmonicity and the dressing effect modify the nature and the number of the bound states as well as the coupling with the remaining phonons. As a consequence, a new theory must be performed to characterize the different pathways for the decay of the TVBS.

\begin{center}
{\bf Appendix A: Unitary transformation to remove the intramolecular anharmonicity}
\end{center}

By assuming that the cubic anharmonicity is one order of magnitude greater than the quartic anharmonicity, a perturbative parameter $\lambda$ is introduced so that the vibrational Hamiltonian of the $n$th amide-I mode is written as
\begin{equation}
h_{n}=h_{n}^{(0)}+\lambda V_{n} + \lambda^{2} W_{n} + ....
\eqnum{A1}
\end{equation}
where $h_{n}^{(0)}= \omega_{0}b^{+}_{n}b_{n}$, $V_{n}= \gamma_{3}(b^{+}_{n}+b_{n})^{3}$ and $W_{n}= \gamma_{4}(b^{+}_{n}+b_{n})^{4}$.
The anharmonic terms in Eq.(A1) can be removed by performing a perturbative unitary transformation $T_{n}=exp(S_{n})$, where $S_{n}$ is expanded as 
\begin{equation}
S_{n}=\lambda S_{n}^{(1)}+\lambda^{2} S_{n}^{(2)} + ...
\eqnum{A2}
\end{equation}
Under this transformation, the Hamiltonian $h_{n}$ becomes
\begin{eqnarray}
&&\tilde{h}_{n}=h_{n}^{(0)}+\lambda (V+[S_{n}^{(1)},h_{n}^{(0)}])+ \nonumber \\
& &+\lambda^{2}(W+[S_{n}^{(2)},h_{n}^{(0)}]+[S_{n}^{(1)},V]+
\frac{1}{2}[S_{n}^{(1}),[S_{n}^{(1)},h_{n}^{(0)}]]) \nonumber \\
& &+...
\eqnum{A3}
\end{eqnarray} 
Since the required unitary transformation must diagonalize the Hamiltonian, the diagonal terms in Eq.(A3) participate in the diagonalization scheme of the Hamiltonian whereas the non diagonal terms, which must vanish, allow us to determine the $S_{n}$ operator order by order. At the second order with respect to the anharmonic parameters, we obtain
\begin{equation}
S_{n}^{(1)}=\frac{\gamma_{3}}{\omega_{0}}[b_{n}^{+3}-b_{n}^{3}+9(b_{n}^{+2}b_{n}-b_{n}^{+}b_{n}^{2}+b_{n}^{+}-b_{n})]
\eqnum{A4}
\end{equation}
and
\begin{eqnarray}
S_{n}^{(2)}&=&\frac{1}{4} [\frac{\gamma_{4}}{\omega_{0}}+3(\frac{\gamma_{3}}{\omega_{0}})^{2}] [b_{n}^{+4}-b_{n}^{4}]+ \nonumber \\
&&[\frac{\gamma_{4}}{\omega_{0}}-3(\frac{\gamma_{3}}{\omega_{0}})^{2}][2b_{n}^{+3}b_{n}-2b_{n}^{+}b_{n}^{3}+3b_{n}^{+2}-3b_{n}^{2}]
\eqnum{A5}
\end{eqnarray}
which lead to the renormalized Hamiltonian 
\begin{equation}
\tilde{h}_{n}=(\omega_{0}-2A)b^{+}_{n}b_{n}- Ab^{+}_{n}b^{+}_{n}b_{n}b_{n}
\eqnum{A6}
\end{equation}
where the irrelevant constant was disregarded and where $A$ denotes the positive anharmonic parameter defined in Eq.(\ref{eq:A}). 

At this step, the full vibron-phonon Hamiltonian $H$ can be expressed in an improved way
by applying the general unitary transformation $T=\prod_{n}T_{n}$. The transformation modifies the lateral interaction as well as the vibron-phonon coupling Hamiltonian leading to the occurrence of vibron-conserving terms and vibron-nonconserving terms. Nevertheless, since the internal frequency $\omega_{0}$ is more than two orders of magnitude greater than the anharmonic parameters $\gamma_{3}$ and $\gamma_{4}$, the nonconserving terms are weak when compared with the conserving terms and will be neglected. Such a  procedure requires the knowledge of the transformation of the vibron displacement operator $b_{n}+b_{n}^{+}$ expressed as
\begin{eqnarray}
& &\tilde{b}_{n}^{+}+\tilde{b}_{n}=b_{n}^{+}+b_{n}+ 
2\frac{\gamma_{3}}{\omega_{0}}[b_{n}^{+2}+b_{n}^{2}-6b_{n}^{+}b_{n}-3] \nonumber \\
& &+[22(\frac{\gamma_{3}}{\omega_{0}})^{2}-6\frac{\gamma_{4}}{\omega_{0}}][b_{n}^{+}+b_{n}+b_{n}^{+2}b_{n}+b_{n}^{+}b_{n}^{2}] \nonumber \\
&&+[3(\frac{\gamma_{3}}{\omega_{0}})^{2}+\frac{\gamma_{4}}{\omega_{0}}][b_{n}^{+3}+b_{n}^{3}]
\eqnum{A7}
\end{eqnarray}
By using Eq.(A7), the transformation of the operators $(b_{n}^{+}+b_{n})^{2}$ and $(b_{n}^{+}+b_{n})(b_{n+1}^{+}+b_{n+1})$ involved in the vibron-phonon Hamiltonian can be determined easily. By restricting the calculations to the vibron-conserving terms, we thus obtain 
\begin{eqnarray}
&&(\tilde{b}_{n}^{+}+\tilde{b}_{n})^{2}=1+88(\frac{\gamma_{3}}{\omega_{0}})^{2}-12\frac{\gamma_{4}}{\omega_{0}}+2(1+2\eta)b_{n}^{+}b_{n} \nonumber \\
&&+2\eta b_{n}^{+}b_{n}^{+}b_{n}b_{n}+\text{nonconserving terms}
\eqnum{A8}
\end{eqnarray}
where $\eta$ is defined in Eq.(\ref{eq:parameter}), and
\begin{eqnarray}
&&(\tilde{b}_{n}^{+}+\tilde{b}_{n})(\tilde{b}_{n'}^{+}+\tilde{b}_{n'})=72(\frac{\gamma_{3}}{\omega_{0}})^{2}[b_{n}^{+}b_{n}+b_{n'}^{+}b_{n'}] \nonumber \\
&&+[1+44(\frac{\gamma_{3}}{\omega_{0}})^{2}-12\frac{\gamma_{4}}{\omega_{0}}][b_{n}^{+}b_{n'}+b_{n'}^{+}b_{n}] \nonumber \\
&&+[22(\frac{\gamma_{3}}{\omega_{0}})^{2}-12\frac{\gamma_{4}}{\omega_{0}}]b_{n}^{+}(b_{n}^{+}b_{n}+b_{n'}^{+}b_{n'})b_{n'} \nonumber \\
&&+[22(\frac{\gamma_{3}}{\omega_{0}})^{2}-12\frac{\gamma_{4}}{\omega_{0}}]b_{n'}^{+}(b_{n}^{+}b_{n}+b_{n'}^{+}b_{n'})b_{n} \nonumber \\
&&+4(\frac{\gamma_{3}}{\omega_{0}})^{2}[b_{n}^{+2}b_{n'}^{2}+b_{n'}^{+2}b_{n}^{2}] 
+144(\frac{\gamma_{3}}{\omega_{0}})^{2} [b_{n}^{+}b_{n'}^{+}b_{n}b_{n'}] \nonumber \\
&&+\text{nonconserving terms}
\eqnum{A9}
\end{eqnarray}

Finally, by using Eqs.(A7)-(A9), the tranformed Hamiltonian Eq.(\ref{eq:HTILDE}) is obtained straightforwardly.

\begin{center}
{\bf Appendix B: Decimation of the two-vibron Schrodinger equation}
\end{center}

For each $k$ value, the two-vibron Schrodinger equation Eqs. (\ref{eq:schrod11})-(\ref{eq:schrod33}) 
can be reduced by using a decimation procedure \cite{kn:pouthier4,kn:pouthier5}. To proceed, let us rewrite the Schrodinger equation Eqs. (\ref{eq:schrod11})-(\ref{eq:schrod33}) as
\begin{eqnarray}
&&(\lambda+2a)\Psi_{k}(0)=-\sqrt{2}g\Psi_{k}(1)  \eqnum{B1} \\
&&(\lambda+b)\Psi_{k}(1)=-\sqrt{2}g\Psi_{k}(0)-\Psi_{k}(2)  \nonumber \\
&&\lambda\Psi_{k}(m)=-\Psi_{k}(m+1)-\Psi_{k}(m-1) ; m=2,3,4,... \nonumber
\end{eqnarray}
where $\lambda=(\omega-2\hat{\omega}_{0})/\Gamma_{k}$, $a=\hat{A}_{k}/\Gamma_{k}$, $b=\hat{B}/\Gamma_{k}$ and $g=\gamma_{k}/\Gamma_{k}$.

In Eq.(B1), we eliminate the wave functions connected to the 
even sites by substituting their expressions in the Schrodinger equation of the odd sites. Nevertheless we keep unchanged the first two Schrodinger equation associated to the defect sites $m=0$ and $m=1$. We thus obtain a new set of Schrodinger equations for the odd sites only, as
\begin{eqnarray}
&&(2-\lambda^{2}-2(\lambda a+1))\Psi_{k}(0)=\sqrt{2}\lambda g\Psi_{k}(1)  \eqnum{B2} \\
&&(2-\lambda^{2}-(\lambda b+1))\Psi_{k}(1)=-\sqrt{2}\lambda g\Psi_{k}(0)-\Psi_{k}(3)  \nonumber \\
&&(2-\lambda^{2})\Psi_{k}(m)=-\Psi_{k}(m+1)-\Psi_{k}(m-1) ; m=3,5,7,... \nonumber
\end{eqnarray}
Eq. (B2) characterizes the Schrodinger equation of the rescaled lattice with new parameters defining the RG transformation 
\begin{eqnarray}
\lambda^{(1)} &=& 2-\lambda^{2} \nonumber \\
a^{(1)}&=&-\lambda a -1 \nonumber \\
b^{(1)}&=&-\lambda b -1 \nonumber \\
g^{(1)}&=&-\lambda g 
\eqnum{B3}
\end{eqnarray}

The previous decimation procedure allows us to define the critical values of the parameters 
responsible for the occurrence of bound states. Indeed, bound states correspond to a localization of the wave function $\Psi_{k}(m)$ around the sites $m=0$ and $m=1$. Therefore, 
when the decimation is applied recursively $p$ times, the neighboring site of the site $m=1$ is pushed to infinity and we thus obtain an ultimate scaled lattice formed by the two side sites $m=0$ and $m=1$, only. The Schrodinger equation can thus be solved exactly and its two eigenvalues are expressed as
\begin{eqnarray}
\lambda^{(p)}&=&-\frac{2a^{(p)}+b^{(p)}}{2} \nonumber \\
&&\pm \sqrt{(\frac{2a^{(p)}+b^{(p)}}{2})^{2}-2(a^{(p)}b^{(p)}-g^{(p)2})}
\eqnum{B4}
\end{eqnarray}

A TVBS occurs when its frequency is at least equal to the minimum of the TVFS continuum. This 
condition is obtained when $\lambda =\lambda _c=-2$. At the critical point 
$\lambda =\lambda _c$, the scaled values of the
parameter $\lambda $ satisfy $\lambda^{(1)}=...=\lambda^{(p)}=\lambda _c$. 
This parameter $\lambda $ becomes scale invariant and is a fixed point of the
RG transformation. Indeed, when the initial lattice dynamics is at a critical point, the localization length $\xi $ is infinite. No change in the critical parameters accompanies the 
length scaling so that the scaled lattice remains at a critical point.

From Eq.(B3), the scaled values of the parameters at the critical 
point are expressed as
\begin{eqnarray}
\lambda^{(p)} &=& -2 \nonumber \\
a^{(p)}&=&2^{p}(a-1)+1 \nonumber \\
b^{(p)}&=&2^{p}(b -1)+1 \nonumber \\
g^{(p)}&=& 2^{p}g 
\eqnum{B5}
\end{eqnarray}
Combining Eqs.(B5) and (B4) for p tends to infinity leads
to the conditions for the occurrence of a bound states in terms of 
the reduced parameters a,b and g as
\begin{equation}
(a-1)(b-1)=g^{2}
\eqnum{B6}
\end{equation}
Eq.(B6) is equivalent to Eq.(\ref{eq:phase}) with the corresponding definition of a, b and g.

\begin{center}
\textbf{Figure Caption}
\end{center}

Figure 1 : (a) Equivalence between the two-vibron Schrodinger equation and the dynamics of a single fictitious particle moving quantum mechanically on a 2D lattice. (b) For each wave vector $k$, the two-vibron dynamics reduces to a one-dimensional tight-binding problem on a semi-infinite lattice which exhibits two defects (see the text).

Figure 2 : Two-vibron energy spectrum for $E_{B}=8$ cm$^{-1}$, $J=8$ cm$^{-1}$, $\Omega_{c}=100$ cm$^{-1}$, $T=310$ K and for (a) $A=0$ cm$^{-1}$ and (b) $A=10$ cm$^{-1}$. 

Figure 3 : Two-vibron energy spectrum for $E_{B}=10$ cm$^{-1}$, $J=8$ cm$^{-1}$, $\Omega_{c}=100$ cm$^{-1}$, $T=310$ K and for (a) $A=0$ cm$^{-1}$ and (b) $A=10$ cm$^{-1}$. 

Figure 4 : Two-vibron energy spectrum for $E_{B}=14$ cm$^{-1}$, $J=8$ cm$^{-1}$, $\Omega_{c}=100$ cm$^{-1}$, $T=310$ K and for (a) $A=0$ cm$^{-1}$ and (b) $A=10$ cm$^{-1}$.

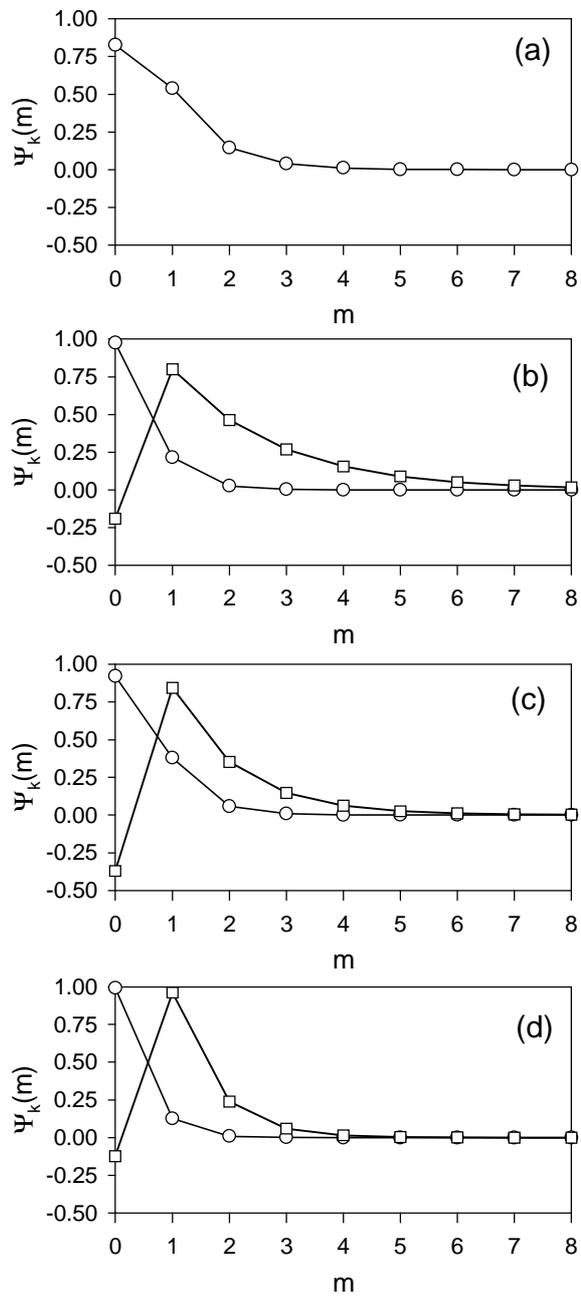
Figure 5 : Zero wave vector two-vibron bound state wave functions for $J=8$ cm$^{-1}$, $\Omega_{c}=100$ cm$^{-1}$, $T=310$ K, and (a) $E_{B}=10$ cm$^{-1}$, $A=0$ cm$^{-1}$, (b) $E_{B}=10$ cm$^{-1}$, $A=10$ cm$^{-1}$, (c) $E_{B}=14$ cm$^{-1}$, $A=0$ cm$^{-1}$, (d) $E_{B}=14$ cm$^{-1}$, $A=10$ cm$^{-1}$. Open circles represent the TVBS-I wave function whereas open squares characterize the TVBS-II wave function (see the text).

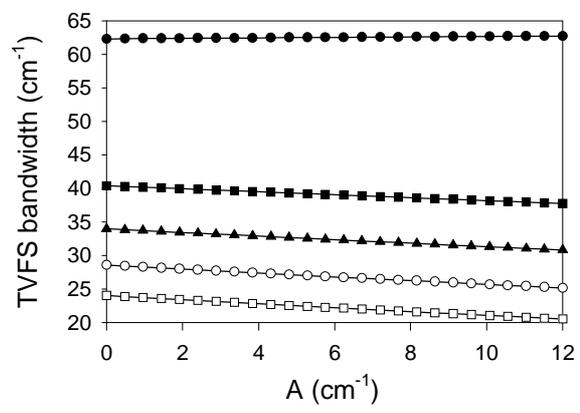
Figure 6 : Behavior of the two-vibron free state bandwidth as a function of the anharmonicity $A$ for $J=8$ cm$^{-1}$, $\Omega_{c}=100$ cm$^{-1}$, $T=310$ K and $E_{B}=0$ cm$^{-1}$ (full circles), $E_{B}=5$ cm$^{-1}$ (full squares), $E_{B}=7$ cm$^{-1}$ (full triangles), $E_{B}=9$ cm$^{-1}$ (open circles), and $E_{B}=11$ cm$^{-1}$ (open squares).

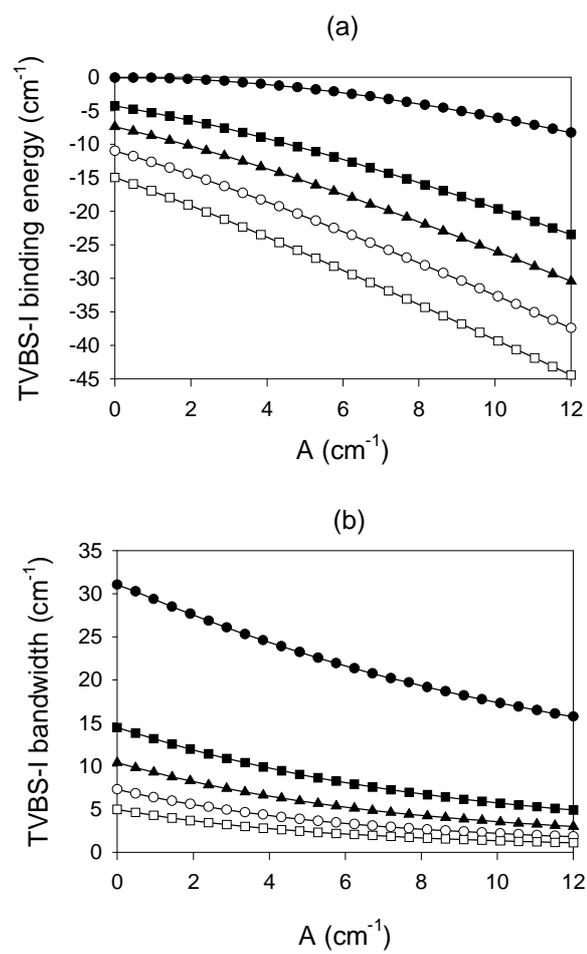
Figure 7 : (a) TVBS-I binding energy and (b) bandwidth as a function of the anharmonicity $A$ for $E_{B}=0$ cm$^{-1}$ (full circles), $E_{B}=5$ cm$^{-1}$ (full squares), $E_{B}=7$ cm$^{-1}$ (full triangles), $E_{B}=9$ cm$^{-1}$ (open circles), $E_{B}=11$ cm$^{-1}$ (open squares) and for $J=8$ cm$^{-1}$, $\Omega_{c}=100$ cm$^{-1}$ and $T=310$ K.

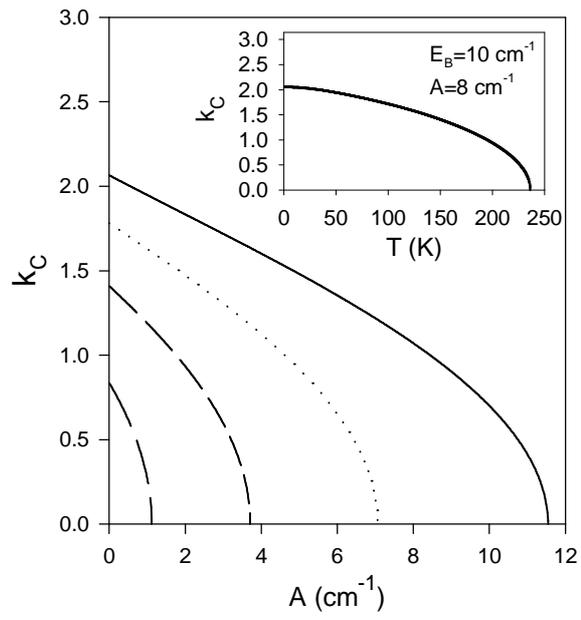
Figure 8 : Behavior of the critical wave vector for which the disappearance of the TVBS-II band takes place as a function of the anharmonicity $A$ for $E_{B}=8$ cm$^{-1}$ (full line), $E_{B}=9$ cm$^{-1}$ (dotted line), $E_{B}=10$ cm$^{-1}$ (short dashed line), $E_{B}=11$ cm$^{-1}$ (medium dashed line) ($J=8$ cm$^{-1}$, $\Omega_{c}=100$ cm$^{-1}$, $T=310$ K). The inset represents the evolution of the critical wave vector as a function of the temperature.

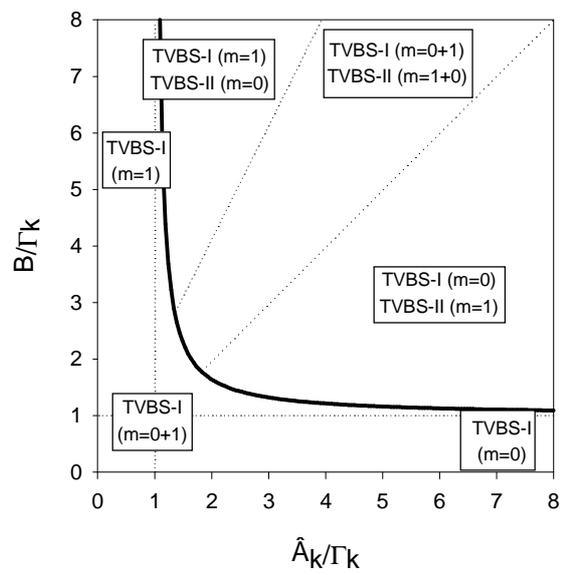
Figure 9 : Phase diagram in the parameter space. The critical curve discriminates a phase with a single bound state from a phase with two bound states (see the text).  

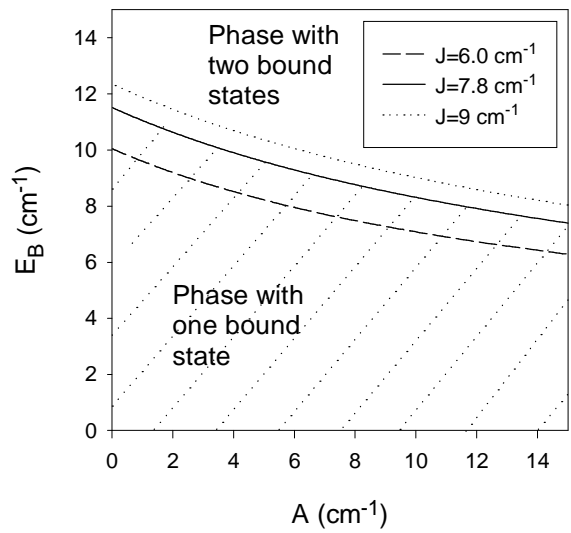
Figure 10 : Phase diagram in the $(A,E_{B})$ parameter space for a zero wave vector ($\Omega_{c}=100$ cm$^{-1}$, $T=310$ K). The critical curve discriminates a phase with a single bound state from a phase with two bound states (see the text).  

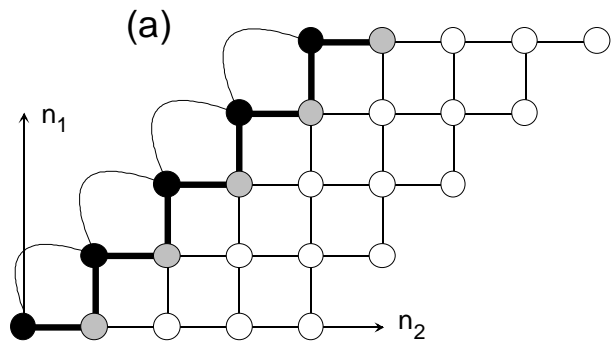
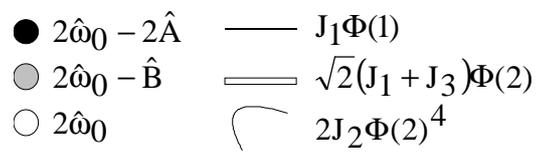
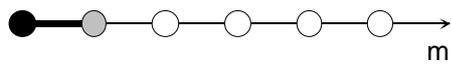
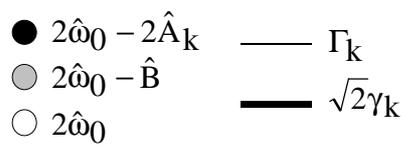
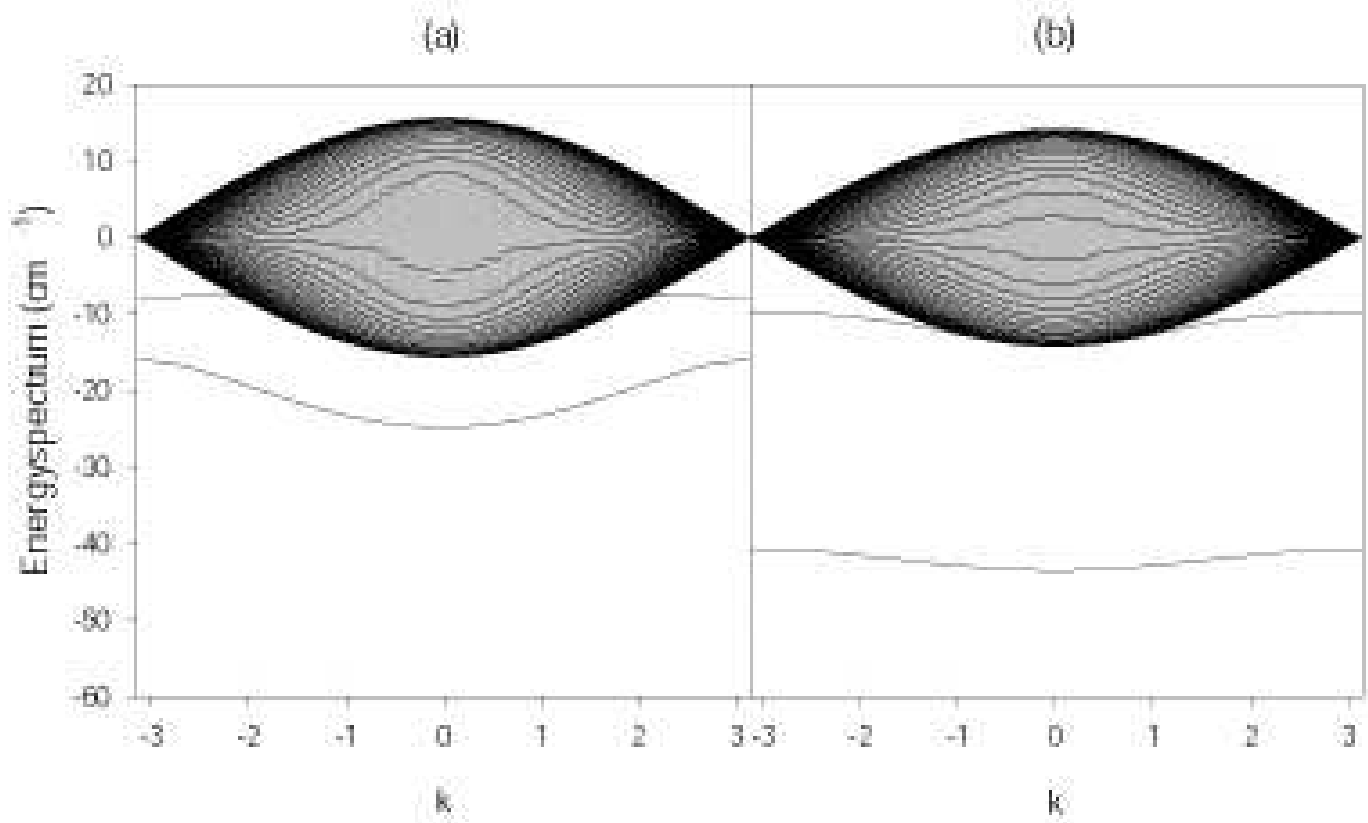
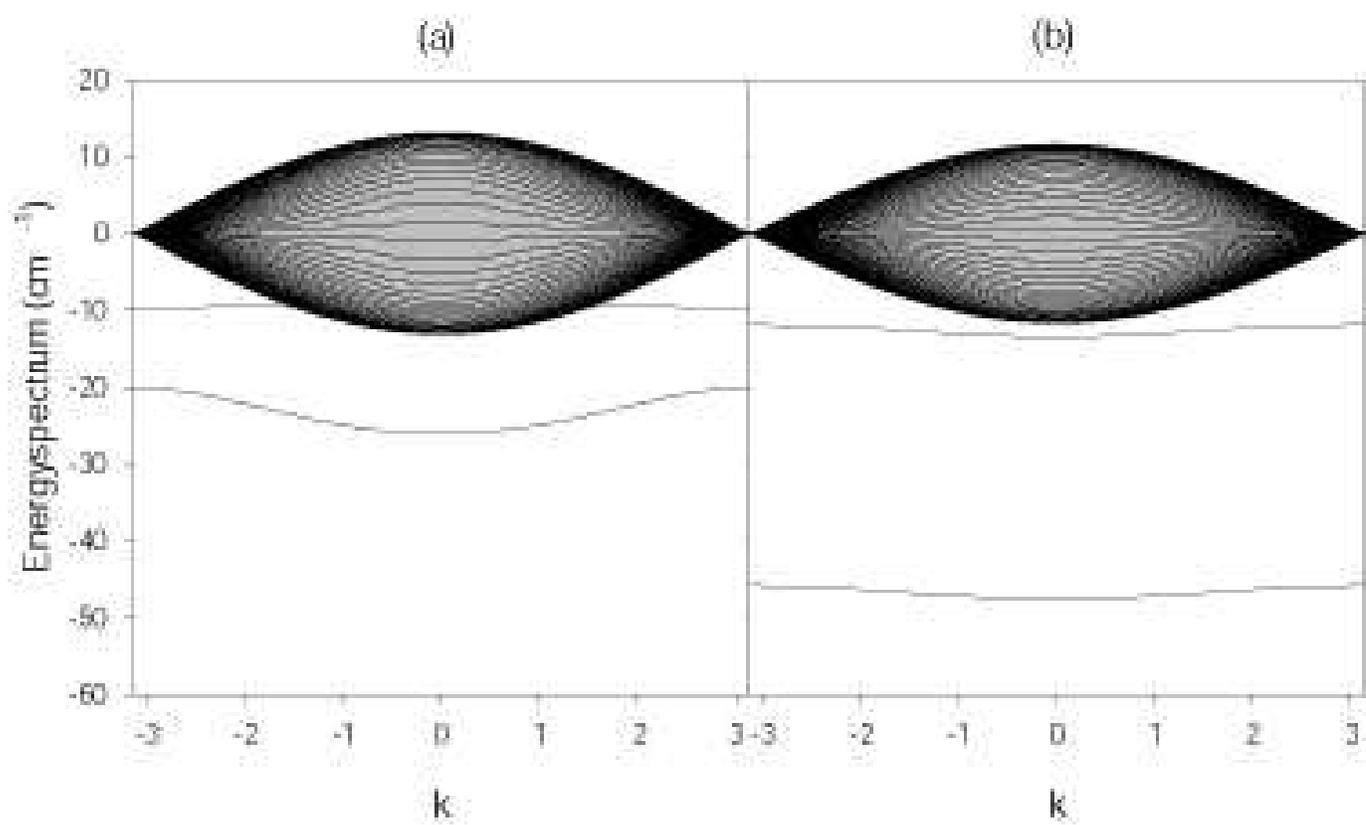
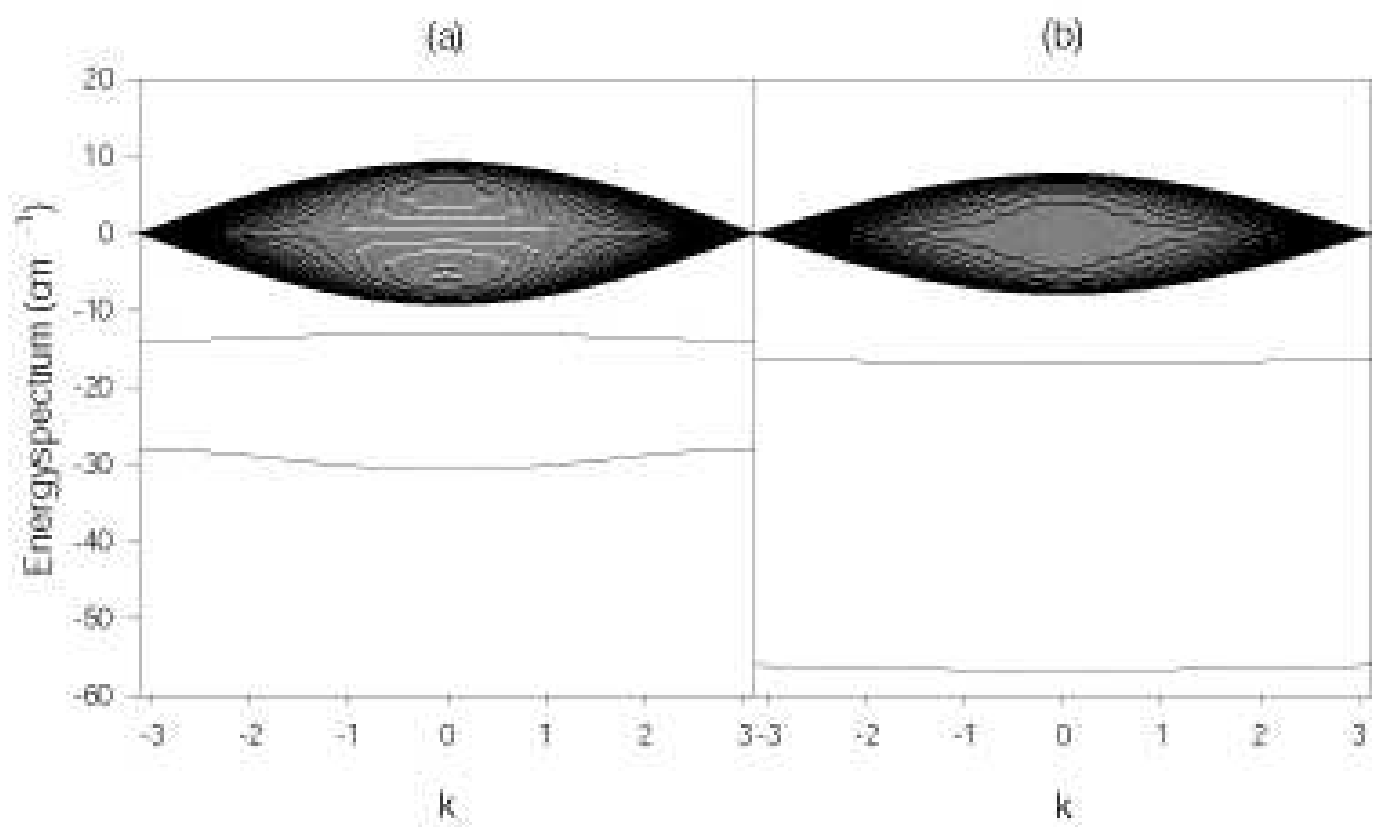
\end{document}